\documentclass[preprint,12pt]{elsarticle}

\usepackage{amssymb}

\usepackage[%
breaklinks=true,
colorlinks=true,
linkcolor=blue,
urlcolor=blue,
citecolor=blue
]{hyperref}

\usepackage{listings}
\usepackage{color}
\usepackage{xcolor}
\definecolor{red}{rgb}{0.6,0,0}
\definecolor{blue}{rgb}{0,0,0.6}
\definecolor{green}{rgb}{0,0.8,0}
\definecolor{cyan}{rgb}{0.0,0.6,0.6}

\usepackage{caption}
\DeclareCaptionFont{white}{\color{white}}
\DeclareCaptionFormat{listing}{\colorbox{blue}{\parbox{\textwidth}{\hspace{15pt}#1#2#3}}}
\usepackage{bm}
\usepackage{amsmath}

\journal{...}

\begin{document}
\definecolor{dkgreen}{rgb}{0,0.6,0}
\definecolor{gray}{rgb}{0.5,0.5,0.5}
\definecolor{mauve}{rgb}{0.58,0,0.82}
\lstset{
  language=Octave,                
  numbers=left,                   
  numberstyle=\tiny\color{gray},     
  stepnumber=2,                   
  numbersep=5pt,
  backgroundcolor=\color{white},      
  showspaces=false,
  showstringspaces=false,
  showtabs=false,
  frame=single,                    
  rulecolor=\color{black},
  tabsize=2,                       
  breaklines=true,                 
  breakatwhitespace=true,          
  title=\lstname,                   
  keywordstyle=\color{blue},          
  commentstyle=\color{dkgreen},       
  stringstyle=\color{mauve}          
}

\begin{frontmatter}



\title{Information system for analysis of nanostructure morphology: Education and research}


\author[label1]{Ludia T. Khusainova}
\author[label1]{Konstantin S. Kolegov}

\affiliation[label1]{organization={Astrakhan Tatishchev State University},
            addressline={20a Tatischev Str.},
            city={Astrakhan},
            postcode={414056},
            country={Russia}}

\begin{abstract}
The work is devoted to the development of an information system \emph{ISANM} that will be useful for teaching students and applicable in the work of engineers and researchers for automating the analysis of colloidal structure morphology. The \emph{ISANM} system integrates various analysis methods and is designed prioritize usability, catering primarily to users rather than software developers. This paper outlines the scope of the subject area, reviews selected methods for morphology analysis, and presents the scripts developed for implementing these methods in \emph{Python}. In the future, new methods and tools will be gradually added to the system for chemical engineers, physicists, and materials scientists. We hope that the system implementation methodology described here will be useful in the implementation of other projects related to training chemical engineers and beyond.
The system core is implemented in \emph{C\#}, utilizing the \emph{.NET Framework} and the \emph{MS SQL Server}, and is developed within the \emph{Microsoft Visual Studio 2019} environment on \emph{Windows 10} using the client-server architecture. The system has been deployed on a \emph{VDS} server using the \emph{Ubuntu 20.04} distribution. \emph{MS SQL Server} is used as the database management system. The system can be accessed at \href{https://isanm.space}{https://isanm.space}. Geometric analysis of colloidal structures is significant in applications such as photonic crystals for optoelectronics and microelectronics, functional coatings in materials science, and biosensors for medical and environmental applications. This demonstrates the importance of digitalization in this area to improve the quality of student education.
\end{abstract}



\begin{keyword}
virtual laboratories \sep information systems \sep micro- and nanoparticles \sep morphology of structures \sep analysis methods \sep databases \sep simulation tools in education



\end{keyword}

\end{frontmatter}


\section*{Introduction}
\label{sec:Introduction}
Digitalization is one of the most important issues in chemical engineering education~\cite{Udugama2023,Barashkin2023}. An example of a 5-year curriculum is given in one pioneering work~\cite{Favre2008}. It includes many training modules, including ``Advances in colloids and interfaces''. The development of digital tools in this field is crucial for the future development of educational and research methods.

Colloidal structures represent both theoretical and practical interest across various fields. They serve as essential building blocks for nanostructured materials with diverse applications, ranging from optoelectronics to the detection of chemical and biological analytes, or the development of biomimetic surfaces with specialized optical or wettability characteristics~\cite{Lotito2017}. The geometry of colloidal structures is closely tied to their physical properties, which holds true not only for ordered periodic structures (notably hexagonal close-packed architectures) that are widely used in practical applications~\cite{Lotito2017}, but also for cases, where specific imperfections in periodic arrangements can affect physical properties~\cite{Chen2013}. Additionally, aperiodic and amorphous patterns are often of interest due to their unique optical and vibrational characteristics~\cite{Lohr2014,Romanov2016,Gratale2016}. Optical, elastic, and wetting properties are strongly dependent on the morphology of colloidal assemblies~\cite{Lotito2020}.

Some existing software solutions support various methods for analyzing colloidal morphology. For instance, \emph{ImageJ} and \emph{Image-Pro Plus} are used for image analysis and processing, while \emph{MATLAB} is utilized for data processing. Another example, \emph{MATBOX} is an open-source toolbox for the commercial software package \emph{MATLAB}~\cite{UsseglioViretta2022}. \emph{MATBOX} is the application for microstructure numerical analysis, including microstructure numerical generation, image filtering and microstructure segmentation, microstructure characterization, three-dimensional visualization, microstructure parameters correlation, and microstructure meshing. The \emph{JavaPlex} library implements persistent homology and related techniques from the field of computational and applied topology. Generally, such software is more oriented towards experienced users or programmers. One of the advantages of \emph{ISANM} is the integration of various analysis methods within a single, accessible service, that does not require local installation. Data storage and processing occur on the server, and only basic computer skills are necessary, making it accessible to engineers and scientists working with colloidal structures and their applications. In addition, the \emph{ISANM} system does not compete with the mentioned programs, such as \emph{ImageJ}. Rather, they complement each other and can be used together. In the future, we will think about integration with different tools. Geometric analysis of colloidal assemblies is important in areas such as photonic crystals for opto- and microelectronics, functional coatings in materials science, and the creation of biosensors for medical and environmental applications. This demonstrates the importance of digitalization in this area to improve the quality of student education.

The purpose of this work is to demonstrate the possibility of creating new digital tools for educational and research activities using \emph{ISANM} development as an example. Here, we share our experience in this area, describe the methodology for implementing such a project, consider the theoretical foundations for analyzing the morphology of colloidal structures, and give instructions on how to use the implemented functionality in practice, which may be useful for educators, students, and researchers.

\section{Data model and methods}
\label{sec:MaterialsAndMethods}
\subsection{Custom Database}
The system uses the \emph{MS SQL Server 2022} for data storage and management. \emph{MS SQL Server 2022} provides enhanced data reliability, scalability, and security, enabling efficient processing and storage of large datasets. All tables have been created based on models using the \emph{Entity Framework} approach.

\begin{figure}[h]
\center{\includegraphics[width=0.99\linewidth]{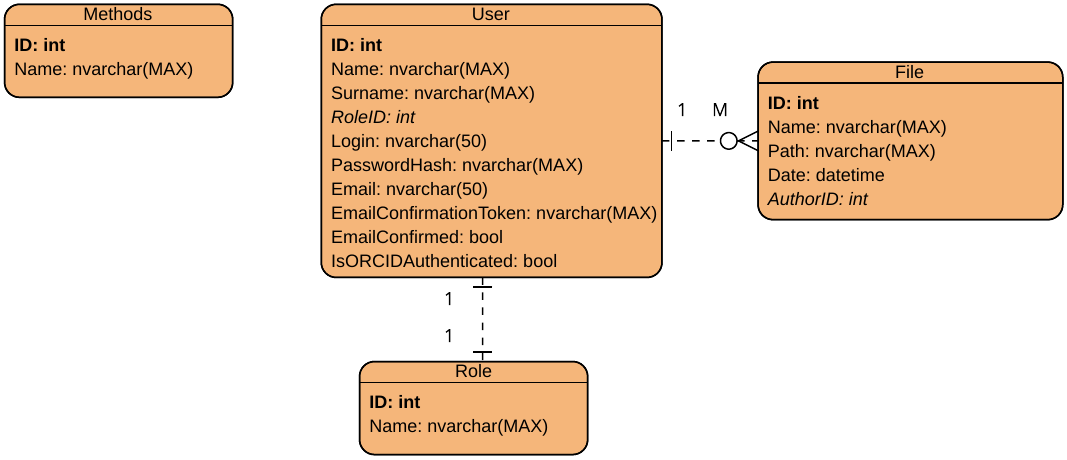}}
\caption{ER diagram.}
\label{fig:ERdiagram}
\end{figure}

The data model of the system comprises the following entities: users, roles, files, and methods. Figure~\ref{fig:ERdiagram} shows the ER diagram illustrating the relationships between these entities.

The ``Users'' entity stores information about all users of the system. This entity is connected by a one-to-one relationship with the ``Roles'' entity. The description of the entity parameters is presented in Table~\ref{tab:UserReference}.

\begin{table}
	\caption{``Users'' entity.}
	\centering
\begin{tabular}{|p{0.35\linewidth}|p{0.15\linewidth}|p{0.15\linewidth}|p{0.15\linewidth}|}
  \hline
  \textbf{Parameter} & \textbf{Type} & \textbf{Size} & \textbf{Range} \\
  \hline
  First name & Text & Maximum & -- \\
  Last name & Text & Maximum & -- \\
  Role & Number & 10 & -- \\
  Login & Text & Maximum & -- \\
  Password hash & Text & Maximum & -- \\
  Email & Text & 50 & -- \\
  Email confirmation token & Text & Maximum & -- \\
  Email confirmed & Boolean & Maximum & True/False \\
  ORCID Authenticated & Boolean & Maximum & True/False \\
  \hline
\end{tabular}
	\label{tab:UserReference}
\end{table}

\begin{table}[h!]
	\caption{``Roles'' entity.}
	\centering
\begin{tabular}{|p{0.2\linewidth}|p{0.2\linewidth}|p{0.2\linewidth}|p{0.2\linewidth}|}
  \hline
  \textbf{Parameter} & \textbf{Type} & \textbf{Size} & \textbf{Range} \\
  \hline
  Name & Text & Maximum & -- \\
  \hline
\end{tabular}
	\label{tab:Roles}
\end{table}

The ``Roles'' entity stores information about user roles. The description of the entity parameters is presented in Table~\ref{tab:Roles}.

The ``Files'' entity stores information about the user files they have uploaded. This entity is connected by a one-to-one relationship with the ``Users'' entity. The description of the entity parameters is presented in Table~\ref{tab:File}.

\begin{table}
	\caption{``Files'' entity.}
	\centering
\begin{tabular}{|p{0.2\linewidth}|p{0.2\linewidth}|p{0.2\linewidth}|p{0.2\linewidth}|}
  \hline
  \textbf{Parameter} & \textbf{Type} & \textbf{Size} & \textbf{Range} \\
  \hline
  Name & Text & Maximum & -- \\
  Path & Text & Maximum & -- \\
  User ID & Number & 10 & -- \\
  Date & Date & 8 & dd.mm.yyyy \\
  \hline
\end{tabular}
	\label{tab:File}
\end{table}

The ``Methods'' entity stores information about methods presented in the system. The description of the entity parameters is presented in Table~\ref{tab:Methods}. The input information is either generated automatically by the system (for example, particle coordinates) or uploaded from the user folder in the system.

\begin{table}
	\caption{``Methods'' entity.}
	\centering
\begin{tabular}{|p{0.2\linewidth}|p{0.2\linewidth}|p{0.2\linewidth}|p{0.2\linewidth}|}
  \hline
  \textbf{Parameter} & \textbf{Type} & \textbf{Size} & \textbf{Range} \\
  \hline
  Name & Text & Maximum & -- \\
  \hline
\end{tabular}
	\label{tab:Methods}
\end{table}

\subsection{Orientational order parameter}
\label{subsec:OrientationOrderParameter}
The orientational order parameter is a tool widely used in condensed matter physics to characterize the local structure of neighboring particles around a specific central particle~\cite{Steinhardt1983,Lotito2019}. It is applicable in various fields, including studies of glasses, jamming transitions, melting, crystallization transitions, and cluster formation. This parameter quantifies the degree of local structural order within a system, reflecting the arrangement of neighboring particles around a given particle. For the analysis of the angular uniformity, the bond orientational order parameter $\psi_{N_{nnj}}^{(j)}$ with respect to an isolated particle $j$ is computed as follows~\cite{Lotito2019}
$$\psi_{N_{nnj}}^{(j)} = \frac{1}{N_{nnj}} \sum_{k=1}^{N_{nnj}} \exp(N_{nnj} i \phi_{jk}),$$
where $\phi_{jk}$ is the angle of the line between the center of the isolated particle and the $k$-th of its nearest neighbors with respect to an arbitrary fixed reference axis (for example, $Ox$), $N_{nnj}$ is the number of its nearest neighbors (coordination number), and $i$ is the imaginary unit.

\begin{figure}[h]
\begin{minipage}[h]{0.49\linewidth}
\center{\includegraphics[width=0.85\linewidth]{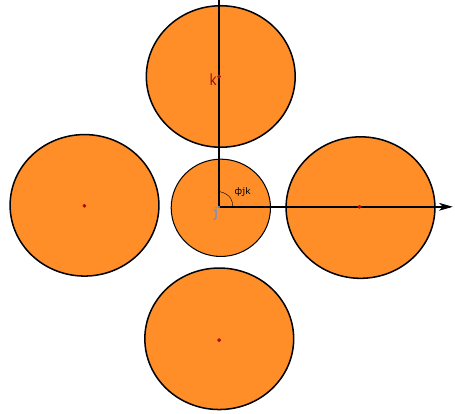} (a)}
\end{minipage}
\hfill
\begin{minipage}[h]{0.49\linewidth}
\center{\includegraphics[width=0.85\linewidth]{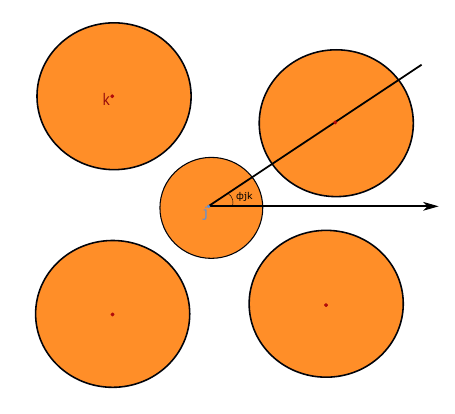} (b)}
\end{minipage}
\caption{Examples of architectures: (a) ideal and (b) chaotic.}
\label{fig:Architecture}
\end{figure}

The closer the value of this parameter is to one, the better the architecture of the insulating particles will be (Fig.~\ref{fig:Architecture}a). Therefore, the closer the value is to zero, the more chaotic the architecture is observed (Fig.~\ref{fig:Architecture}b).

\begin{figure}[h]
\begin{minipage}[h]{0.49\linewidth}
\center{\includegraphics[width=0.85\linewidth]{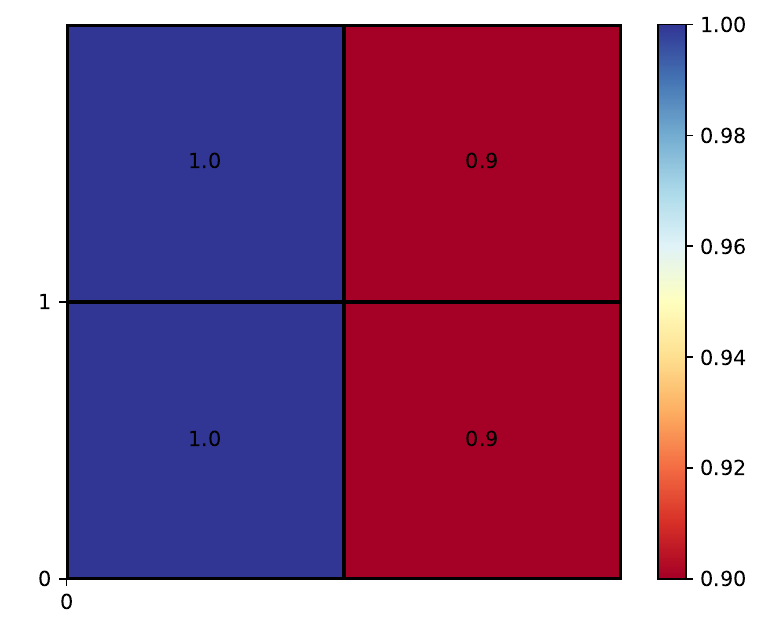} (a)}
\end{minipage}
\hfill
\begin{minipage}[h]{0.49\linewidth}
\center{\includegraphics[width=0.75\linewidth]{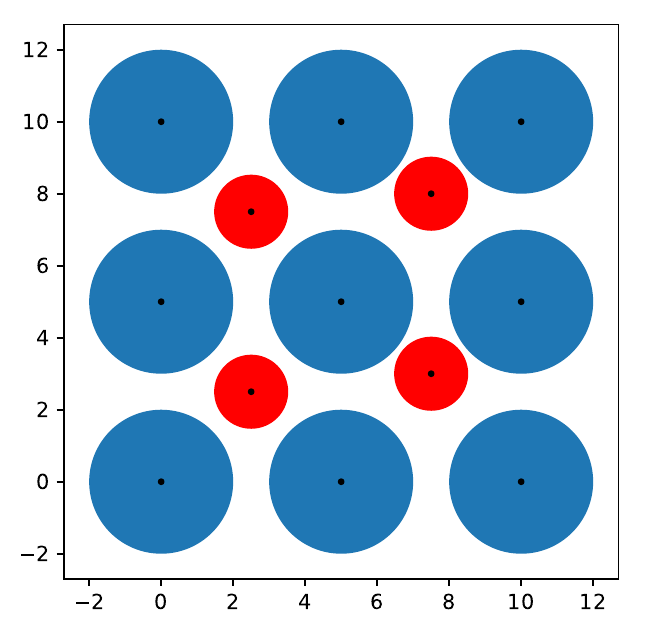} (b)}
\end{minipage}
\caption{(a) Heatmap for the parameter $\psi_{N_{nnj}}^{(j)}$ and (b) particle visualization.}
\label{fig:HeatmapParticleVisualization}
\end{figure}

In the implemented method, the plane is divided into subregions (squares). The number of subregions is specified manually, while their width is determined automatically. The user inputs the number of subregions along both the $Ox$ and $Oy$ axes. It is also necessary to define the particle status in the binary mixture (i.e., large and small particles), specifying which particle sizes are considered primary and which are auxiliary. The parameter $\psi_{N_{nnj}}$ is calculated for each primary isolated particle, and then averaged over all primary isolated particles within each subregion. Thus, the value of $\left\langle \psi_{N_{nnj}} \right\rangle$ is calculated for each subregion. Based on this data, a heatmap is generated (Fig.~\ref{fig:HeatmapParticleVisualization}a).

\begin{figure}[h]
\begin{minipage}[h]{0.49\linewidth}
\center{\includegraphics[width=0.75\linewidth]{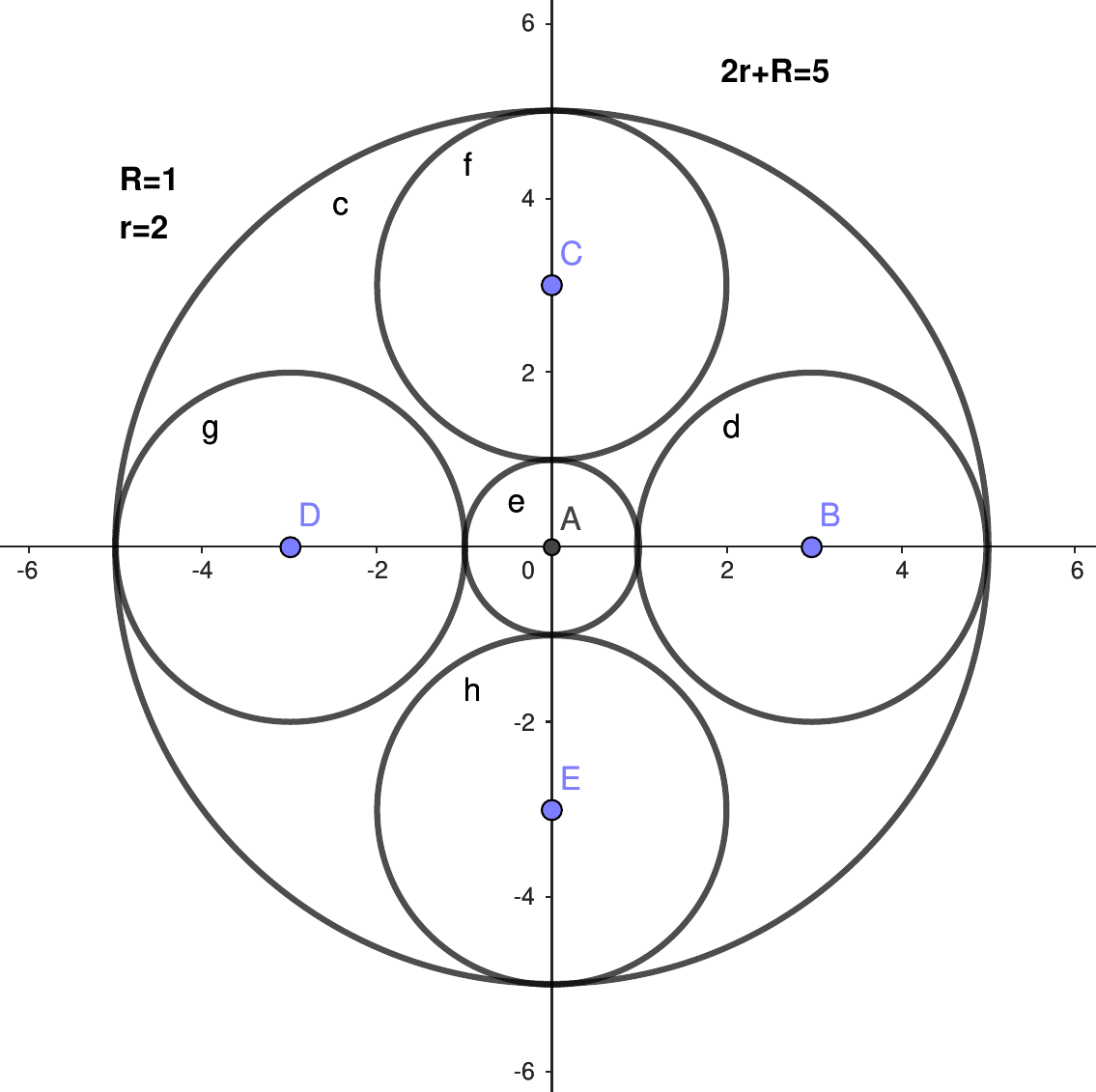} (a)}
\end{minipage}
\hfill
\begin{minipage}[h]{0.49\linewidth}
\center{\includegraphics[width=0.78\linewidth]{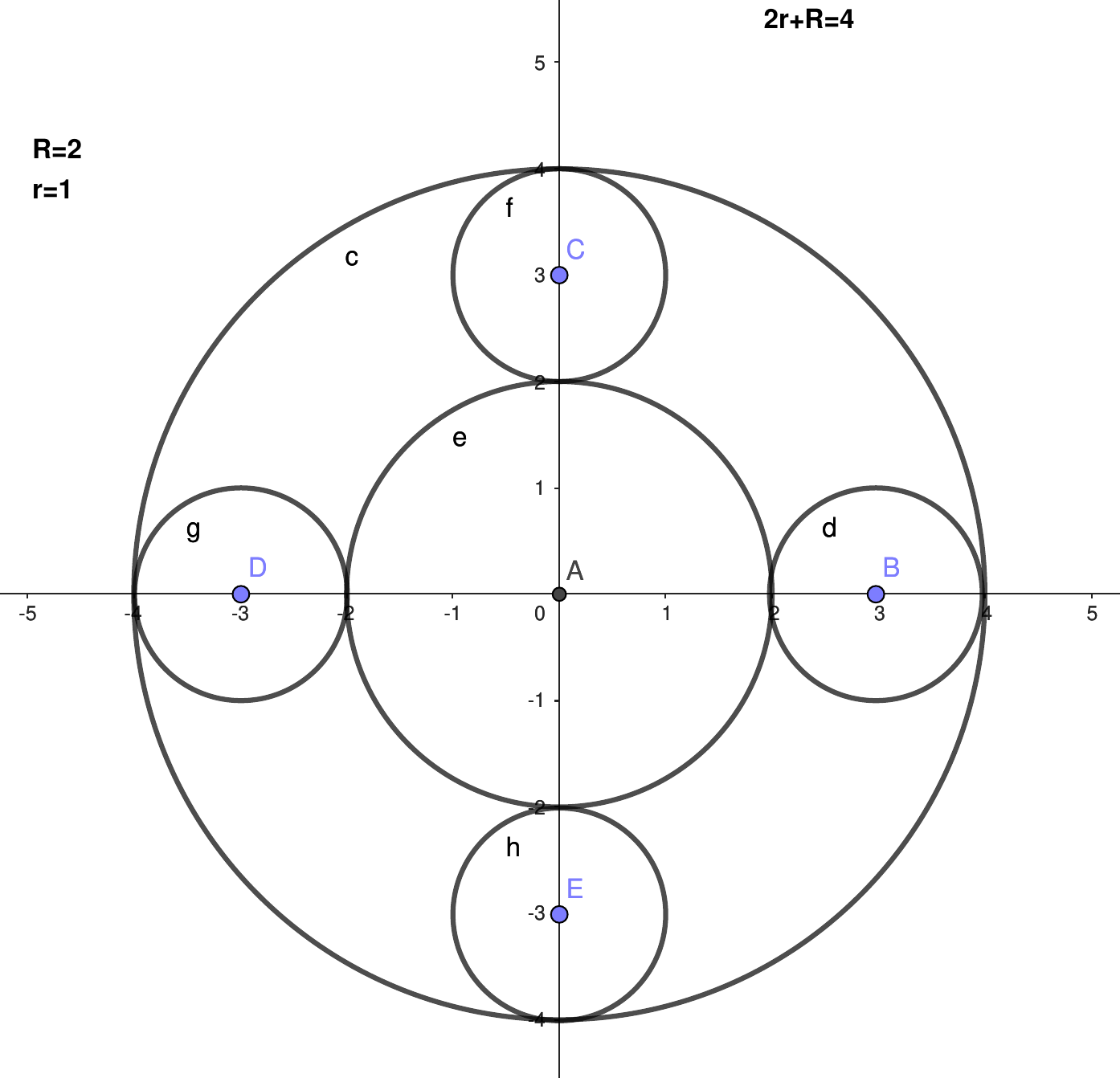} (b)}
\end{minipage}
\caption{(a) Definition of the nearest L neighbors and (b) S neighbors.}
\label{fig:NearestNeighborsDetermination}
\end{figure}

To determine the number of nearest neighbors, the radii of both the primary and auxiliary particles are taken into account. The radius of the circle within which the nearest neighbors must be located is calculated as $2r + R$, where $R$ is the radius of the primary particle and $r$ is the radius of the auxiliary particle (Fig.~\ref{fig:NearestNeighborsDetermination}a). The center of this imaginary circle is aligned with the center of each primary particle for which it is constructed. It is important to note that this implementation does not take into account dimers of isolated particles. Only individual isolated particles are considered.

A user should provide a \emph{.txt} file containing the coordinates of particles (Fig.~\ref{fig:HeatmapParticleVisualization}b). There are two options available: either upload an existing file, or generate a new one automatically by specifying the desired number of particles. For example, the user selects a file located in their folder. Next, the type of the main particle must be manually selected. After clicking the ``Run'' button, the script starts processing the input data and subsequently provides the results (Fig.~\ref{fig:HeatmapParticleVisualization}a), which can then be saved.

\subsection{Radial uniformity}
\label{subsec:RadialUniformity}

Radial uniformity describes the property of a material or system that tends to exhibit uniform characteristics in the radial direction relative to a central point, axis, or center of mass. This concept can be applied in various fields, such as physics, engineering, or geometry. In the context of colloidal structures, radial uniformity may refer to the uniform distribution of colloidal particles relative to the center of the colloidal system. Colloidal systems include particles ranging in size from a few nanometers to micrometers, dispersed within a medium such as water or another liquid. Assessment of the radial uniformity of colloidal particles may include measurement and analysis of particle concentrations in various radial zones or distances from the center of a colloidal system.

To assess radial uniformity~\cite{Lotito2019}, the following parameter was used $$C_{jN_{jSL}}= \frac{1}{N_{nnjSL}} \sum_{k=1}^{N_{nnjSL}} \frac{\left| t_{jk} - \bar t_j \right|}{\bar t_j},$$ where $t_{jk}$ is the distance between the center of the L particle $j$ and the $k$-th S particle among its nearest neighbors, $N_{nnjSL}$ is the number of neighbors, $\bar t_j$ is the average distance from the center of the L particle to the S particles among $N_{nnjSL}$ neighbors.

\begin{figure}[h]
\begin{minipage}[h]{0.49\linewidth}
\center{\includegraphics[width=0.85\linewidth]{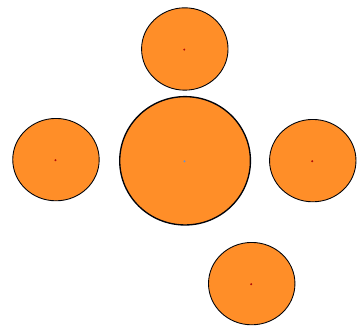} (a)}
\end{minipage}
\hfill
\begin{minipage}[h]{0.49\linewidth}
\center{\includegraphics[width=0.85\linewidth]{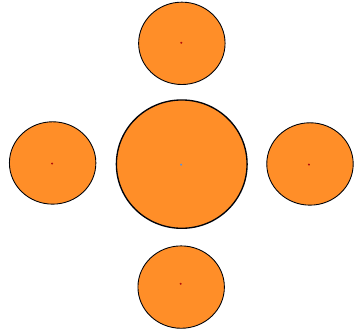} (b)}
\end{minipage}
\caption{(a) Uneven and (b) uniform particle distribution.}
\label{fig:UnevenAndUniformParticleDistribution}
\end{figure}

The key aspect of this parameter is the variation in distances between the isolated L particle and the surrounding S particles. If all surrounding particles are equidistant from the isolated one (Fig.~\ref{fig:UnevenAndUniformParticleDistribution}b), indicating a uniform distribution of particles, then the value of the parameter $C_{jN_{jSL}}$ is zero. Thus, with a less uniform distribution (Fig.~\ref{fig:UnevenAndUniformParticleDistribution}a), the value of this parameter will tend toward one.

\begin{figure}[h]
\begin{minipage}[h]{0.49\linewidth}
\center{\includegraphics[width=0.85\linewidth]{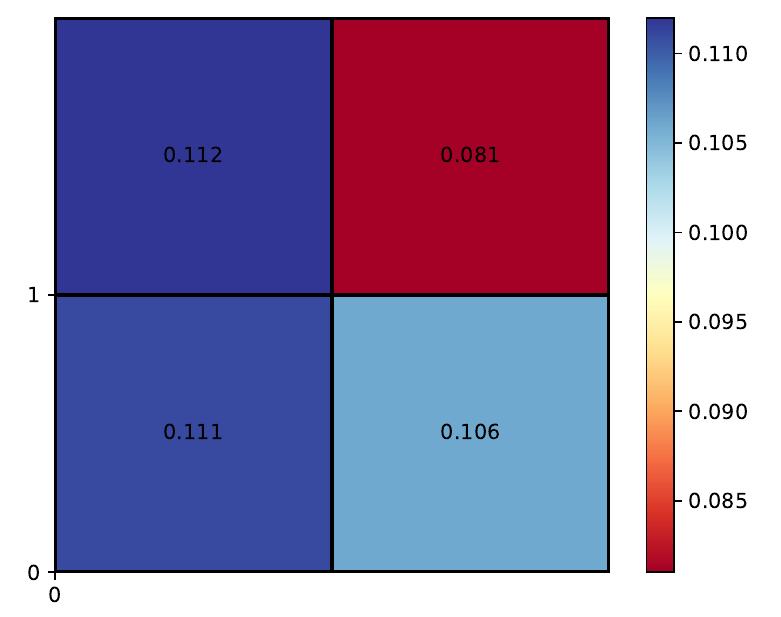} (a)}
\end{minipage}
\hfill
\begin{minipage}[h]{0.49\linewidth}
\center{\includegraphics[width=0.85\linewidth]{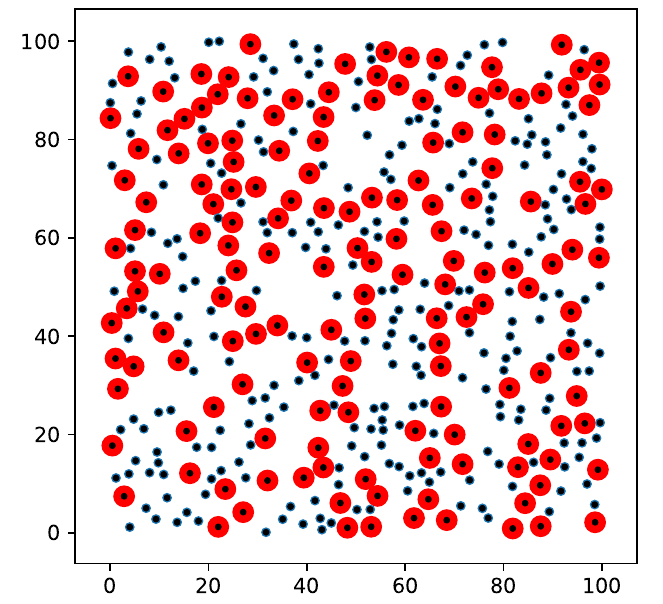} (b)}
\end{minipage}
\caption{(a) Heatmap for the $C_{jN_{jSL}}$ parameter and (b) visualization of particles.}
\label{fig:HeatmapParticleVisualization2}
\end{figure}

In the implemented method, the plane is divided into subregions (squares). The subdivision into subregions is performed similarly to the ``Orientational order parameter'' method (see subsection~\ref{subsec:OrientationOrderParameter}). The parameter $C_{jN_{jSL}}$ is calculated for each primary isolated L particle (Fig.~\ref{fig:HeatmapParticleVisualization2}b), and then averaged over all primary isolated L particles within each subregion. Heatmap (Fig.~\ref{fig:HeatmapParticleVisualization2}a) is constructed, based on this data.

To determine the number of nearest neighbors, the radii of both the primary and auxiliary particles are taken into account. The radius of the circle within which the nearest neighbors must be located is calculated as $2r + R$, where $R$ is the radius of the primary particle and $r$ is the radius of the auxiliary particle (Fig.~\ref{fig:NearestNeighborsDetermination}b). The center of this imaginary circle is aligned with the center of each primary particle for which it is calculated. It is important to note that this implementation does not take into account dimers of isolated particles. Only individual isolated particles are considered.

The user should provide a \emph{.txt} file containing the coordinates of particles. There are two options available: either upload an existing file, or generate a new one automatically by specifying the desired number of particles. For example, the user automatically generates 420 particles arranged in a disordered manner (Fig.~\ref{fig:HeatmapParticleVisualization2}b). The status of the primary particle is set automatically, as this method is applied only to the surrounding S particles around the L particle. After clicking the ``Run'' button, the script starts processing the input data and subsequently provides the results (Fig.~\ref{fig:HeatmapParticleVisualization2}a), which can later be saved.

\subsection{Bond length}

Bond length is a complex topic that requires an understanding of both colloids and the concept of bond length. A colloid is a mixture in which one substance, composed of microscopically dispersed insoluble particles, is suspended in another substance. The particles of the dispersed phase have diameters ranging from approximately 1~nm to 1~$\mu$m. In this context, bond length is the average distance between the nuclei of two bonded atoms in a molecule. Bond length is a property of the bond between fixed-type atoms that is relatively independent of the rest of the molecule. This means that bond length depends on the types of atoms forming the bond, rather than the environment in which the bond exists.

\begin{figure}[h!]
\center{\includegraphics[width=0.7\linewidth]{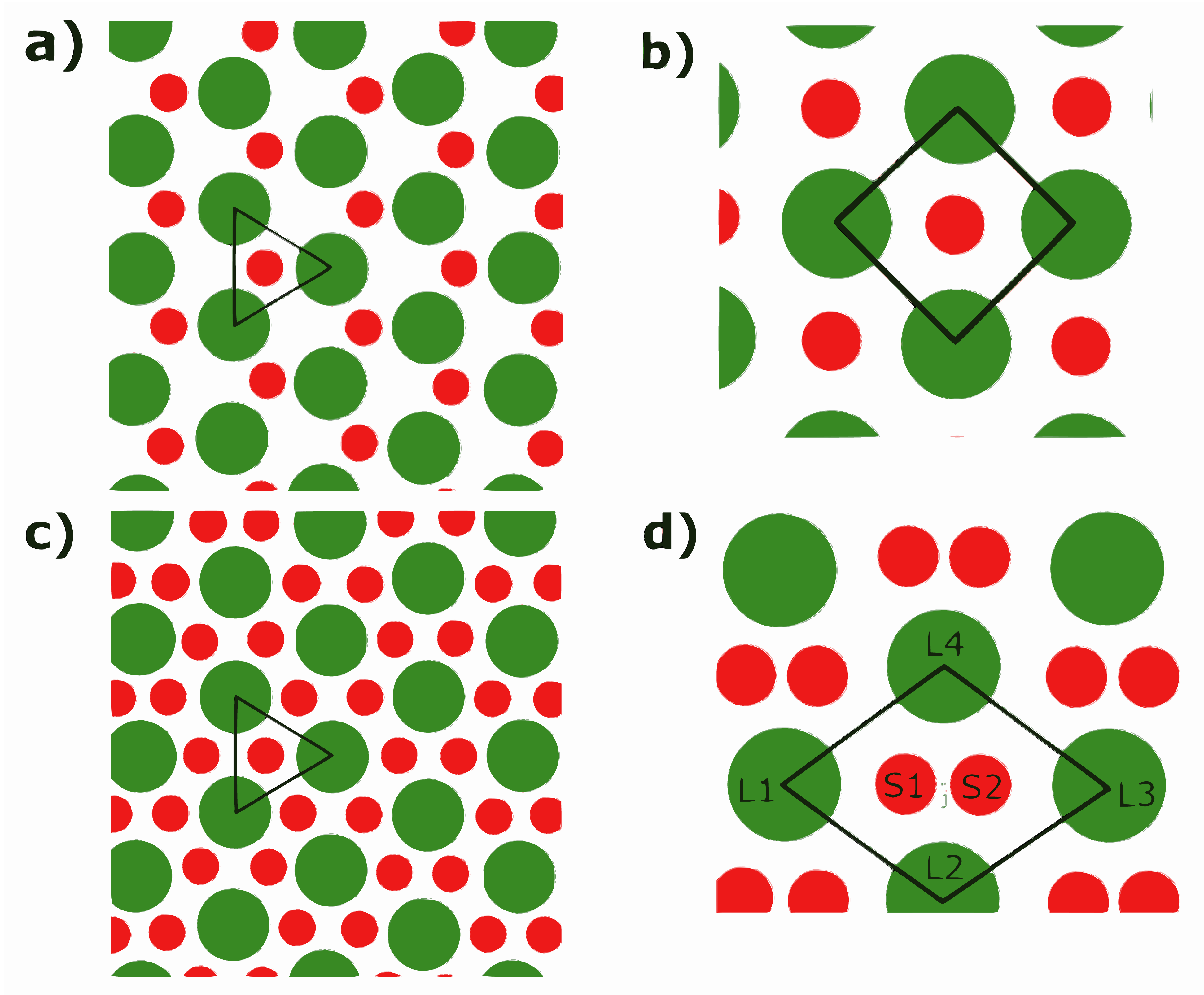}}
\caption{Configuration of L particles around S particles: a) isolated S particles, present alternatively in the interstices between L particles arranged in a hexagonal pattern L particles; b) isolated S particles present in the interstices between L particles arranged in a square pattern; c) isolated S particles, present in all the interstices between L particles arranged in a hexagonal pattern; d) S particle dimers present between L particles arranged in a rhombic pattern.}
\label{fig:ConfigurationOfLargeParticles}
\end{figure}

In the context of colloidal structures, it is important to note that bond length may vary depending on the type of particles and the medium in which they are suspended. For example, Figure~\ref{fig:ConfigurationOfLargeParticles} shows the configurations of L particles around an isolated S particle or a dimer of S particles.

There is a configuration with the potential for periodic space filling, namely, an isolated S particle situated between four L particles, located at the vertices of square. This configuration is distinguished by examining two different metrics: the orientational order parameter $\psi_{4LS}^{(j)}$ (see subsection~\ref{subsec:OrientationOrderParameter}) and bond length $b_{jNLS}$, defined as~\cite{Lotito2019} $$b_{jNLS} = \frac{1}{N} \sum_{k=1}^{N} \frac{\left| l_{jk} - \bar l_j \right|}{\bar l_j},$$ where $l_{jk}$ is the distance between the S particle $j$ and the nearest neighbor L particle $k$, and $\bar l_j$ is the average distance between the S particle $j$ and the $N$ L particles, with $N = 4$ for the square arrangement. For a perfect square pattern, $\left| \psi_{4LS}^j \right| = 1$ and $b_{j4LS} = 0$.

The key point is the change in distances between the isolated S particle and the surrounding L particles. If all the surrounding particles are equidistant from the isolated one, meaning the distribution of particles is uniform, the value of the parameter $b_{jNLS}$ equals zero. Consequently, with a less uniform distribution, the value of this parameter will approach one.

There is a generalized procedure to identify all the different configurations. Such a procedure is based on the analysis of the angular and radial uniformity (see subsections~\ref{subsec:OrientationOrderParameter} and \ref{subsec:RadialUniformity}) of L particles around S particle or around the middle point of an S particle dimer. First, L nearest neighbors are determined via Voronoi tessellation/Delaunay triangulation. Subsequently, for isolated S particles and the middle points of S particle dimer placed within the convex quadrilateral generated by four L neighbor particles, the angular and radial distributions of L particle neighbors with respect to the isolated S particle or to the middle point of the S particle dimer are scrutinized~\cite{Lotito2019}.

In the implemented method, the plane is divided into subregions (squares). The subdivision into subregions is performed similarly to the ``Orientational order parameter'' method (see subsection~\ref{subsec:OrientationOrderParameter}). The parameter $b_{jNLS}$ is calculated for each isolated S particle (Fig.~\ref{fig:HeatmapParticleVisualization2}b). The value of this parameter is averaged over all primary isolated S particles within the subregion. Thus, the value of $b_{jNLS}$ is determined for each subregion. Heatmap is constructed, based on this data. The number of nearest neighbors is determined in the same way as in the radial uniformity estimation method (see subsection~\ref{subsec:RadialUniformity}).

\begin{figure}[h]
\center{\includegraphics[width=0.4\linewidth]{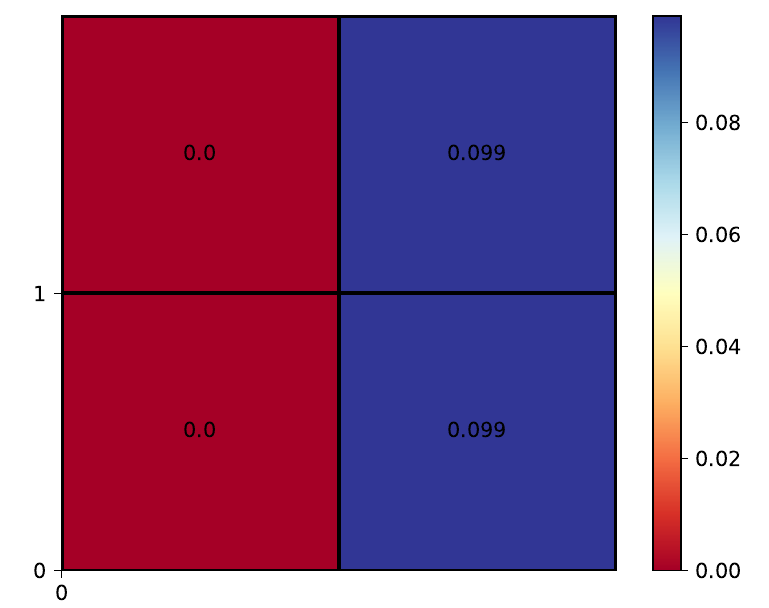}}
\caption{Heatmap for the parameter $b_{jNLS}$.}
\label{fig:HeatmapBondLength}
\end{figure}

The user should provide a \emph{.txt} file containing the particle coordinates. There are two options available: either upload an existing file, or generate a new one automatically by specifying the desired number of particles. For example, the user provides an existing file located in their folder as input. The status of the primary particle is set automatically, as this method is applied only to the surrounding L particles around the S particle. After clicking the ``Run'' button, script starts processing the input data (Fig.~\ref{fig:HeatmapParticleVisualization}b) and subsequently provides the results (Fig.~\ref{fig:HeatmapBondLength}), which can later be saved.

\subsection{Voronoi diagram}

A Voronoi diagram is a partition of a plane into regions, each of which is closest to one of a given set of objects. In the simplest case, these objects are a finite number of points on the plane, referred to as seed points, nodes, or generators. To construct the Voronoi diagram, each cell can be built sequentially.

There is a simple algorithm for constructing the Voronoi diagram. For each point on the plane, the intersection of its circle with the circles centered around all other points must be computed. This intersection forms the boundary of the Voronoi region for that point. However, computing the intersection for each point has a time complexity of $O(n^4)$, where $n$ is the number of points on the plane. Another algorithm for constructing the Voronoi diagram is a recursive algorithm, which divides the plane into four parts and then recursively continues dividing each part until each region contains only one point. This algorithm has a time complexity of $O(n \log n)$.

The Voronoi diagram has wide applications in various fields, such as computer graphics, geometric modeling, and optimization. For example, it is used to solve the Nelson--Erd{\"o}s--Hadwiger problem, which involves finding an upper bound for the chromatic number of a Euclidean space of dimension 2 or 3. The properties of the Voronoi diagrams are closely related to the Delaunay triangulations, where Delaunay triangulation is a partition of a plane into triangles with no two points lying inside the circumcircles of any triangle. There is a one-to-one correspondence between Voronoi diagrams and Delaunay triangulations. If one connects points whose Voronoi regions share boundaries with edges, one gets the Delaunay triangulation graph.

In colloidal systems, the Voronoi diagram can be used to determine the size and shape of colloidal particles. Colloidal particles can be defined as the centers of Voronoi cells, where the volume of the dispersed phase penetrates into the cell. Since the Voronoi cells correspond to the nearest points, the particle sizes can be determined by the distance between the centers of neighboring cells.

\begin{lstlisting}[float,caption={Processing coordinates from file},label=lst:CoordinateProcessing]
    # reading coordinates from a file for a Voronoi diagram
with open(Filename, 'r') as f:
    for lines in f:
        l1 = lines[0:-3]
        l2 = l1.split('\t')
        l2[x] = float(l2[x])
        l2[y] = float(l2[y])
        l = []
        l.append(l2[x])
        l.append(l2[y])
        # defining the processing area
        if l[x]>x_bot_border and l[x]<x_up_border and l[y]>y_bot_border and l[y]<y_up_border:
            points.append(l)
    f.close()
\end{lstlisting}

In the implemented method, the area for constructing the diagram is set automatically. The particle coordinates and the configuration file are uploaded manually by the user, and data processing is carried out (Listings~\ref{lst:CoordinateProcessing} and \ref{lst:ProcessingConfigurationFile}).

\begin{lstlisting}[float,caption={Processing the configuration file},label=lst:ProcessingConfigurationFile]
    # processing and assigning the variables from the array
for i in range(len(variables)):
    if variables[i][0] == 'Number of particles':
        Np = float(variables[i][1])
    if variables[i][0] == 'radius of particles_1':
        rp1 = float(variables[i][1])
        dp1 = 2 * rp1
    if variables[i][0] == 'radius of particles_2':
        rp2 = float(variables[i][1])
        dp2 = 2 * rp2
    if variables[i][0] == 'L':
        scaleLen = float(variables[i][1])
\end{lstlisting}

Next, using the \emph{scipy} library, a Voronoi diagram is constructed for the points bounded by the specified area (Listing~\ref{lst:VoronoiDiagramGenerationFunction}). It is worth noting the input parameters of the \emph{voronoi\_plot\_2d} function.

\begin{lstlisting}[float,caption={Voronoi diagram construction function},label=lst:VoronoiDiagramGenerationFunction]
# Creating Voronoi diagram.
# vor - A diagram to create (scipy.spatial.Voronoi instance).
# show_vertices - Adding coordinate vertices to a plot (bool, optional).
# show_points - Adding Voronoi points to the graph (bool, optional).
# line_colors - line color (string, optional).
# line_width - line width (float, optional).
# point_size - point size (float, optional).
vor = Voronoi(points)
fig = voronoi_plot_2d(vor, show_vertices=0, show_points=0, line_colors='black', line_width=0.7, point_size=1)
\end{lstlisting}

After that, the particles are rendered on the Voronoi diagram according to their radius (Listing~\ref{lst:ParticleRendering}). The coordinates of the particles vary in the range from $-1$ to $1$.

\begin{lstlisting}[float,caption={Particle rendering},label=lst:ParticleRendering]
for i in range(len(points_for_dots)):
    # checking the particle radius
    if points_for_dots[i][r]<(rp1+rp1*0.01) and points_for_dots[i][r]>(rp1-rp1*0.01):
        circle1 = plt.Circle((points_for_dots[i][x], points_for_dots[i][y]), rp1/scaleLen, color='blue')
        plt.gcf().gca().add_artist(circle1)
    if points_for_dots[i][r]<(rp2+rp2*0.01) and points_for_dots[i][r]>(rp2-rp2*0.01):
        circle2 = plt.Circle((points_for_dots[i][x], points_for_dots[i][y]), rp2/scaleLen, color='brown')
        plt.gcf().gca().add_artist(circle2)
\end{lstlisting}

\begin{figure}[h!]
\center{\includegraphics[width=0.5\linewidth]{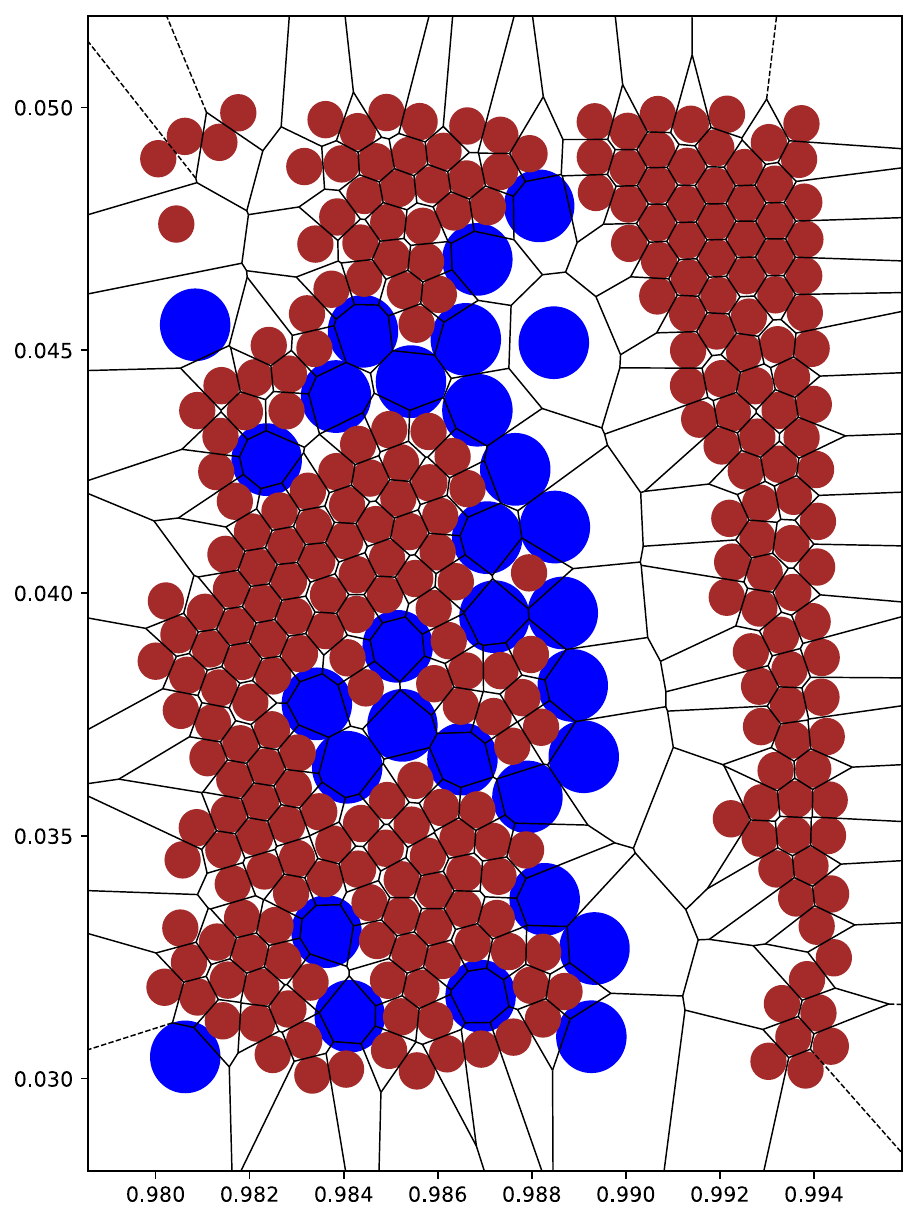}}
\caption{Voronoi diagram for a binary mixture of particles~\cite{Zolotarev2022} (\copyright 2022 AIP Publishing).}
\label{fig:VoronoiDiagramForBinaryMixtureOfParticles}
\end{figure}

The user provides two \emph{.txt} files as input: one file contains particle coordinates and the other is a configuration file with details about the experiment. After clicking the ``Run'' button, the script begins to execute, and the user is presented with a Voronoi diagram as output (Fig.~\ref{fig:VoronoiDiagramForBinaryMixtureOfParticles}).

\subsection{Mean squared displacement}

In statistical mechanics, the Mean Squared Displacement (MSD) is a measure of the deviation of a particle's position from its initial position over time. It is commonly used to study the motion of particles in various physical and biological systems. The MSD function provides insight into the dynamics and behavior of particles within a given system. This dynamic is related to the mechanical properties of the medium through which the particles move. MSD can be used to analyze the diffusion properties of particles, determine the nature of their motion (e.g., normal or anomalous), and understand the relationship between particle movement and system parameters, such as temperature or microstructure. The MSD function can be derived using various mathematical approaches, such as quantum mechanics or Lévy random walks, depending on the specific system under investigation.

MSD at a given time is defined as the ensemble average (statistical mechanics)~\cite{Zolotarev2021}
$$\langle l^2 \rangle =\frac{1}{N_p} \sum_{i=1}^{N_p}\left( \left( x_i(t) - x_i(0) \right)^2 + \left( y_i(t) - y_i(0) \right )^2 \right),$$ where $N_p$ is the number of particles and $t$ is the time.

Einstein and Smoluchowski developed a mathematical framework describing the Brownian motion of particles, establishing a quantitative relationship between the mean displacement of a particle (mean squared displacement, displacement amplitude) and the diffusion coefficient $D$. The relationship they derived between these two quantities is known as the Einstein--Smoluchowski equation. During its motion, a molecule collides with other molecules that prevent it from moving in a straight line. If we examine the path in more detail, we can see that this approach is a good approximation of random walk. Such walking consists of a series of successive steps, each taken in a completely random direction relative to the previous one. This type of path was analyzed by Albert Einstein in his study of Brownian motion, and he showed that the mean square distance traveled by a particle after random walk is proportional to the time elapsed. In two dimensions, this relationship can be expressed as $\langle l^2 \rangle = 2 d\, D\, t$, where $d$ is the dimensionality of the system ($d = 2$ for the 2D case). However, a linear dependence on time is assumed. Otherwise, the formula would differ slightly, indicating anomalous diffusion. From this, we can derive the formula for the diffusion coefficient $D = \langle l^2 \rangle / (2 d\,  t)$. This formula demonstrates the relationship between MSD and the diffusion coefficient. Furthermore, we can approximately calculate the displacement distance of a particle over a small time step $\Delta t$ as $l = \sqrt{2d\, \Delta t\, D}$.

In the implemented method, the MSD parameter is calculated for each particle. The value of this parameter is averaged over all particles in the experiment. A graph is plotted to show the dependence of the averaged MSD value on the time step. It should be noted that spatial coordinates must be normalized in the input data files. The coordinates ($x$, $y$) vary in the range from $-1$ to $1$.

\begin{figure}[h]
\center{\includegraphics[width=0.7\linewidth]{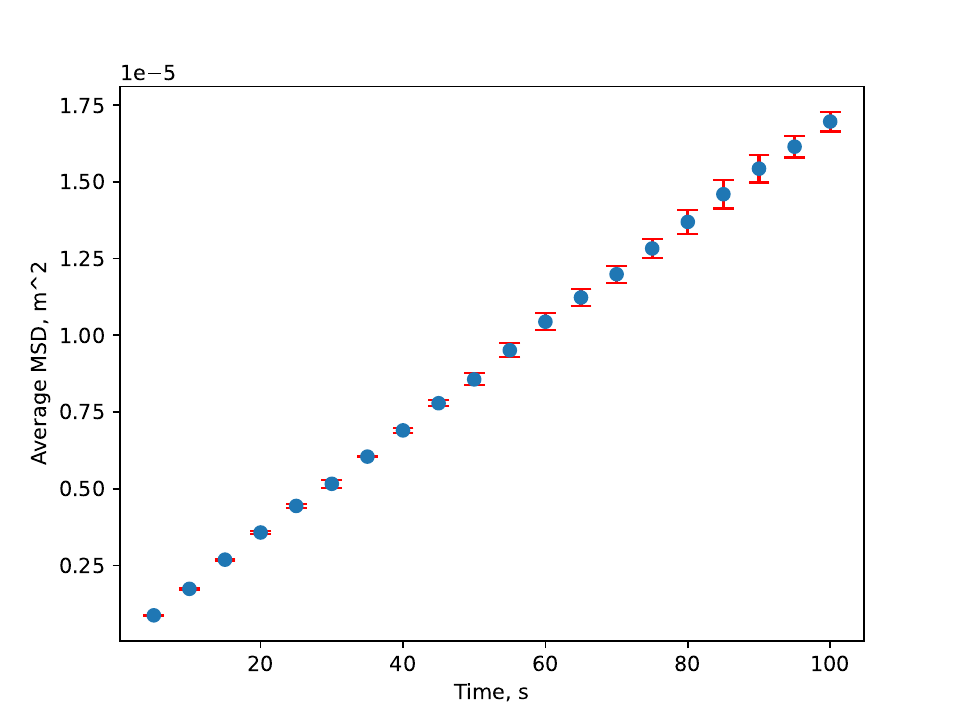}}
\caption{Dependence of MSD on time.}
\label{fig:MSD_vs_time}
\end{figure}

In this method, the user should provide an archive containing numbered folders of experiments (e.g., ten experiments), each of which includes \emph{.txt} files with particle coordinates (for more details, see the method page at \href{https://isanm.space}{https://isanm.space}). The user can automatically compute the error, such as the ``Mean Square Deviation'' or the ``Standard Error of the Mean'', by selecting the appropriate checkbox. Once the ``Run'' button is clicked, the method executes, and the user receives the output (Fig.~\ref{fig:MSD_vs_time}).

\subsection{Concentric tetratic order parameter}
\label{subsec:ConcentricParameterTetraticOrder}

In the experiment~\cite{Sanchez2015}, a system of tubular particles arranged in a cylindrical cell was studied. A specially developed analysis was used to determine the degree of order of such structures. This analysis can also be applied to other geometries with different boundary shapes.

Elongated rigid particles are often encountered in experiments, for example, in granular systems, as well as in certain biological and colloidal systems, such as nanotubes. The morphology of the aggregation of elongated particles can be characterized by orientational order.

The tetratic orientational order effectively characterizes the state of a system when components align along or perpendicular to a basis vector, which can occur in the volume. However, the tangent direction to the boundary in the shape of a circle is not fixed; it depends on the angular coordinate (polar coordinate system). Therefore, it is useful to define parameters that take into account the direction tangent to the nearest point on the boundary of the area in the form of a circle. In the region under consideration, this corresponds to the vector $\boldsymbol{\Phi}$, which is directed from the point $(x_c,\, y_c)$ perpendicular to the vector extending from the center of the coordinate system to the point $(x_c,\, y_c)$. Thus, we define a concentric tetratic order parameter as $S_4 = \left\langle \cos 4\theta \right\rangle$, where $\theta$ is the angle of deviation of the rod from the basis vector $\boldsymbol{\Phi}$, originating from the central point $(x_c,\, y_c)$ and extending tangentially to the circle with radius~\cite{Sanchez2015} $$r = \sqrt{x_c^2 + y_c^2}.$$ The angular brackets denote averaging over all particles that belong to a certain subregion, for example, within a ring of a specified thickness.

\begin{figure}[h]
\center{\includegraphics[width=0.5\linewidth]{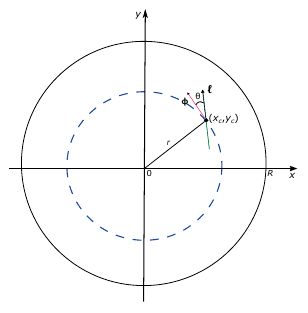}}
\caption{Determining the angle of deviation $\theta$.}
\label{fig:DeterminationOfTheta}
\end{figure}

Suppose the coordinates of the central and edge points of an elongated particle are known. In Fig.~\ref{fig:DeterminationOfTheta} the vector $\mathbf{l}= (x_1 - x_c, y_1 - y_c)$ is directed along the particle from the center to the edge. To determine the coordinates of the vector $\boldsymbol{\Phi}$, we rotate the vector $\mathbf{r}$ by $90^\circ$, while scaling the coordinates by the length of $\mathbf{r}$. When rotating the vector by $90^\circ$, its coordinates swap places, and one of the coordinates is taken with a negative sign, resulting in $$\boldsymbol{\Phi} = \left( \frac{-y_c}{r},\, \frac{x_c}{r} \right).$$

\begin{figure}[h]
\center{\includegraphics[width=0.95\linewidth]{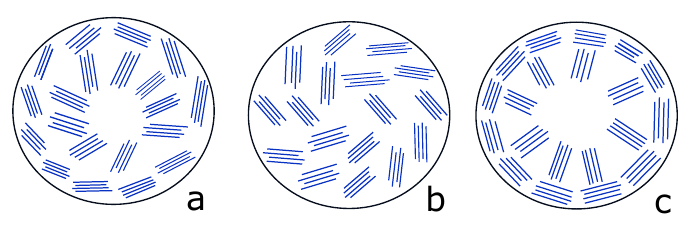}}
\caption{Examples of configurations: (a) $\left\langle S_4 \right\rangle = 0$, (b) $\left\langle S_4 \right\rangle = -1$ and (c) $\left\langle S_4 \right\rangle = 1$.}
\label{fig:ExamplesOfConfigurations}
\end{figure}

This parameter becomes zero for the isotropic state (Fig.\ref{fig:ExamplesOfConfigurations}a). In this case, the value of $\theta$ can be approximately $22.5^\circ$ or $67.5^\circ$. Since the basis vector is determined by the angular position of the tangent, this parameter can also take negative values. Therefore, for the ideal tetratic state (Fig.~\ref{fig:ExamplesOfConfigurations}b) with a basis vector consistently oriented at $45^\circ$ to the particle's orientation, this parameter $S_4 = -1$. The highest value of tetratic order is $S_4 = 1$ (Fig.~\ref{fig:ExamplesOfConfigurations}c). In this case, the value of $\theta$ is $0^\circ$ or $90^\circ$.

We find the angle $\theta$ between two vectors using the formula
\begin{equation}\label{eq:CosTheta}
  \cos \theta = \frac{\boldsymbol{\Phi} \cdot \mathbf{l}}{|\boldsymbol{\Phi}|\,|\mathbf{l}|},
\end{equation}
where $$|\boldsymbol{\Phi}| = \sqrt{\frac{y_c^2}{r^2} + \frac{x_c^2}{r^2}}= 1,\, |\mathbf{l}| = \sqrt{(x_1 - x_c)^2 + (y_1 - y_c)^2},$$ and the scalar product of two vectors is $$\boldsymbol{\Phi} \cdot \mathbf{l} = - \frac{y_c (x_1 -x_c)}{r} + \frac{x_c (y_1 - y_c)}{r}.$$ Therefore, we obtain the formula
\begin{equation}\label{eq:CosTheta2}
  \cos \theta = \frac{- \frac{y_c (x_1 -x_c)}{r} + \frac{x_c (y_1 - y_c)}{r}}{\sqrt{(x_1 - x_c)^2 + (y_1 - y_c)^2}}.
\end{equation}
It follows from the double angle formula that
\begin{multline*}
    \cos 4\theta = \cos^2 2\theta - \sin^2 2\theta = (\cos^2 2\theta - \sin^2 2 \theta)^2 - (2 \sin \theta \cos \theta)^2 =\\
    (\cos^2 \theta - \sin^2 \theta)^2 - 4 \sin^2 \theta \cos^2 \theta =\\
    (\cos^2 \theta - 1 + \cos^2 \theta)^2 - 4 (1 - \cos^2 \theta) \cos^2 \theta =\\
    (2 \cos^2 \theta - 1)^2 - 4 \cos^2 \theta + 4 \cos^4 \theta =\\
    8 \cos^4 \theta - 8 \cos^2 \theta + 1.
\end{multline*}

In the implemented method, the circle is divided into subregions (rings). The number of rings is specified manually, while their width is calculated automatically. It is also possible to set the boundaries for the coordinate $r$, which is included within the range $[0,\, R]$, where calculations will be performed. For each rod, the parameter $S_4$ is calculated. The value of this parameter is averaged over all rods within the subregion. Thus, the value of $S_4$ is determined for each subregion. A graph  is plotted to display the dependence of $S_4$ on the coordinate $r$. It is important to note that spatial coordinates must be normalized with respect to the radius $R$ in the input files. The dimensionless value of the radius of the circle corresponds to one, while the coordinates $x$ and $y$ vary in the range from $-1$ to $1$.

\begin{figure}[h]
\center{\includegraphics[width=0.5\linewidth]{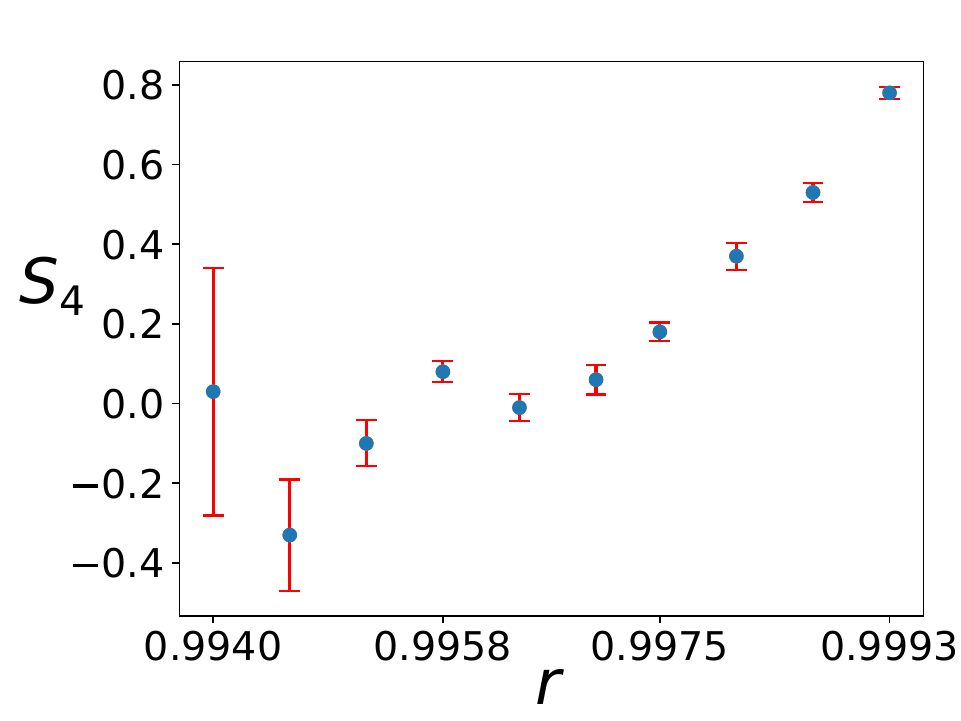}}
\caption{Dependence of the concentric tetratic order parameter on the radius.}
\label{fig:S_4VsRadius}
\end{figure}

In this method, the user should provide an archive containing the particle coordinates (for more details, see the method page at \href{https://isanm.space}{https://isanm.space}). The boundaries for the calculations can be entered manually or automatically determined based on the particle coordinates. It is necessary to specify the number of rings into which the circle will be divided. The user also has the option to calculate the Mean Square Deviation or the Standard Error of the Mean. Once the ``Run'' button is clicked, the method executes, and the user receives the output (Fig.~\ref{fig:S_4VsRadius}).

\subsection{Nematic order parameter}

Elongated rigid particles are frequently encountered in experiments and can be found in granular systems, as well as in certain biological and colloidal systems, such as nanotubes. These elongated particles may exhibit orientational order, typically characterized by the following parameter~\cite{Sanchez2015,Donev2006}
$$S_m = \max\limits_{\theta_0}\left\langle \cos \left[ m (\theta - \theta_0) \right] \right\rangle.$$

For $m = 2$, we obtain a nematic order parameter that equals one for the ideal nematic phase and zero for the isotropic phase. Here, $\theta_0$ corresponds to the direction along which the particles align. If we take the vector $\boldsymbol{\Phi}$ for such a direction and count from it, then $\theta_0 = 0$. In this case, we have $S_2 = \left\langle \cos 2\theta \right\rangle$, where $\theta$ is the angle of deviation of the rod from the basis vector  $\boldsymbol{\Phi}$, originating from the central point $(x_c,\, y_c)$ and extending tangentially to the circle with radius $r$ (Fig.~\ref{fig:DeterminationOfTheta})~\cite{Wang2022}. The angular brackets denote averaging over all particles that belong to a certain subregion, for example, within a ring of a specified thickness.

Assuming the coordinates of the central and edge points of an elongated particle are known. In the Fig.~\ref{fig:DeterminationOfTheta}, the vector $\mathbf{l} = (x_1 - x_c, y_1 - y_c)$ is directed along the particle from the center to the edge (see subsection~\ref{subsec:ConcentricParameterTetraticOrder}). The angle $\theta$ between the two vectors is determined using formula~\eqref{eq:CosTheta}, leading us to formula~\eqref{eq:CosTheta2}. From the double-angle formula, it follows that $$ \cos 2\theta = 2 \cos^2 \theta - 1.$$

In a two-dimensional (2D) system, we obtain $S_2 = 2 \cos^2 \theta - 1$, where $\theta$ is the angle between the optimally aligned direction and the main axis of the particle (the direction along which it is elongated). The nematic order parameter $S_2$ ranges from $-1$ to $1$, with higher absolute values representing greater order. The sign of $S_2$ is positive if nanotubes tend to align parallel to the chosen direction, and negative if they tend to align perpendicular to it~\cite{Zhao2015}.

In the implemented method, the circle is divided into subregions (rings). The number of rings is specified manually, while their width is calculated automatically. It is also possible to set the boundaries for the coordinate $r$, which is included within the range $[0,\, R]$, where calculations will be performed. For each rod, the parameter $S_2$ is calculated. The value of this parameter is averaged over all rods within the subregion. Thus, the value of $S_2$ is determined for each subregion. A graph  is plotted to display the dependence of $S_2$ on the coordinate $r$. It is important to note that spatial coordinates must be normalized with respect to the radius $R$ in the input files. The dimensionless value of the radius of the circle corresponds to one, while the coordinates $(x,\, y)$ vary in the range from $-1$ to $1$.

\begin{figure}[h]
\center{\includegraphics[width=0.5\linewidth]{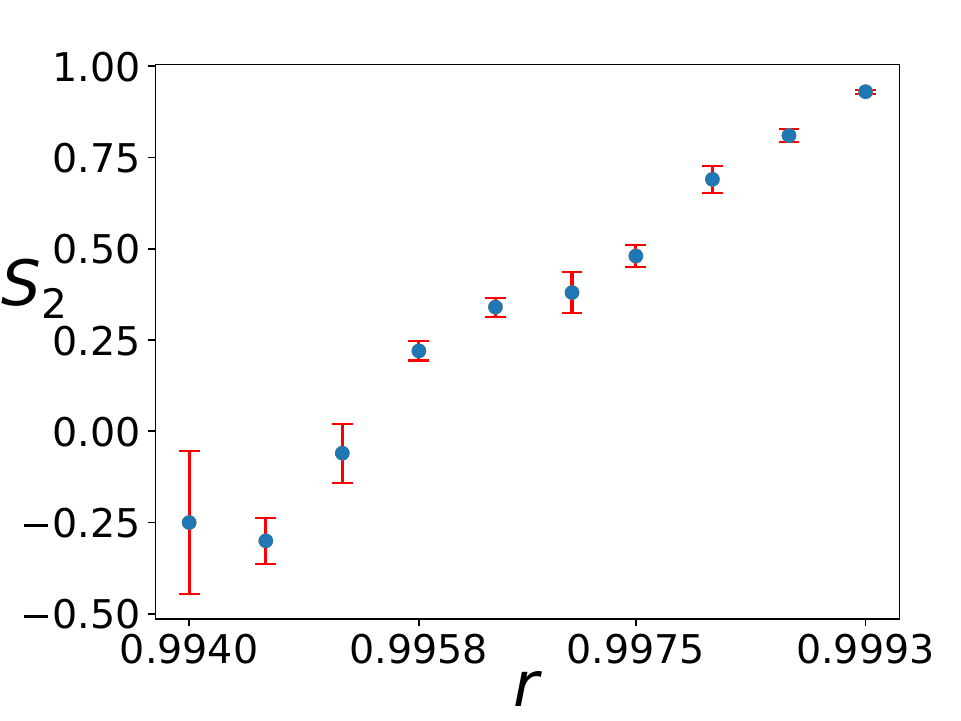}}
\caption{Dependence of the nematic order parameter on the radius.}
\label{fig:S_2VsRadius}
\end{figure}

The practical implementation of this method is similar to the approach described in subsection~\eqref{subsec:ConcentricParameterTetraticOrder}. The distinguishing feature lies in the results obtained from the method execution (Fig.~\ref{fig:S_2VsRadius}).

\subsection{Shape factor}

The shape factor is a dimensionless quantity employed in image analysis and microscopy to characterize the shape of a particle independently of its size. Generally, a colloid is defined as a mixture in which one substance consists of microscopically dispersed insoluble particles suspended in another medium. Some colloids appear translucent due to the Tyndall effect, which occurs when light is scattered by the particles within the colloid. In contrast, other colloids may be opaque or exhibit faint coloration.

The classification of angles $\alpha$ in a polygon formed by connecting L particles surrounding an
S particle can assist determine whether the polygon predominantly exhibits a rhombic or square configuration. This can be achieved by evaluating the ratio of the diagonals of the polygon. Alternatively, the shape factor $\vartheta_{jLS}$ can be assessed to determine the angles $\alpha$ of the formed polygon. The shape factor~\cite{Lotito2019}, in relation to the isolated S particle or the midpoint $j$ of the dimer formed from S particles can be expressed as $$\vartheta_{jLS}=\frac{p_{jLS}^2}{4\pi A_{jLS}},$$ where $p_{jLS}$ and $A_{jLS}$ represent the perimeter and area of the polygon, which has vertices consisting of the four L nearest neighbors to the isolated S particle or the midpoint $j$ of the dimer formed from S particles.

From the definition, it can be concluded that $\vartheta_{jLS} = 1$ corresponds to a circular arrangement, while $\vartheta_{jLS} > 1$ indicates regular polygons. For instance, we have $\vartheta_{jLS} = 1.1563$ for a regular pentagon and $\vartheta_{jLS} = 1.073$ for a regular hexagon~\cite{Lotito2018}.

The area of the formed polygon is calculated using Gauss's area formula
\begin{multline*}
  A_{jLS} = \frac{1}{2} \left| \sum_{i=1}^{n-1} x_i y_{i+1} + x_n y_1 -  \sum_{i=1}^{n-1} x_{i+1} y_i - x_1 y_n \right| =  \\
  = \frac{1}{2} \left| x_1 y_2 + x_2 y_3 + \dots + x_{n-1} y_n + x_n y_1 - x_2 y_1 - x_3 y_2 -  \dots - x_n y_{n-1} - x_1 y_n \right|,
\end{multline*}
where $n$ is the number of sides of the polygon and $(x_i,\, y_i)$ are the coordinates of the polygon's vertices.

In the implemented method, the plane is divided into subregions (squares). This subdivision is performed similarly to the ``Orientational order parameter'' method (see subsection~\ref{subsec:OrientationOrderParameter}). Users can specify the statuses of the particles, indicating which particles are classified as primary and which as auxiliary. For each primary isolated particle, the parameter $\vartheta_{jLS}$ is computed. The value of this parameter value is averaged across all primary isolated particles within a given subregion. Consequently, the value of $\vartheta_{jLS}$ is determined for each subregion. Heatmap is constructed, based on this data. The number of nearest neighbors is determined using the same approach as in the method for calculating the ``Orientational order parameter'' (see subsection~\ref{subsec:OrientationOrderParameter}).

\begin{figure}[h]
\begin{minipage}[h]{0.49\linewidth}
\center{\includegraphics[width=0.8\linewidth]{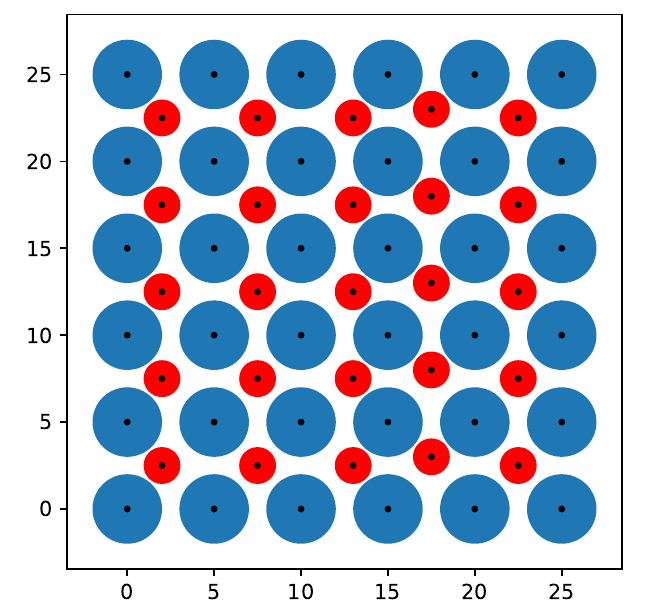} \\(a)}
\end{minipage}
\hfill
\begin{minipage}[h]{0.49\linewidth}
\center{\includegraphics[width=0.95\linewidth]{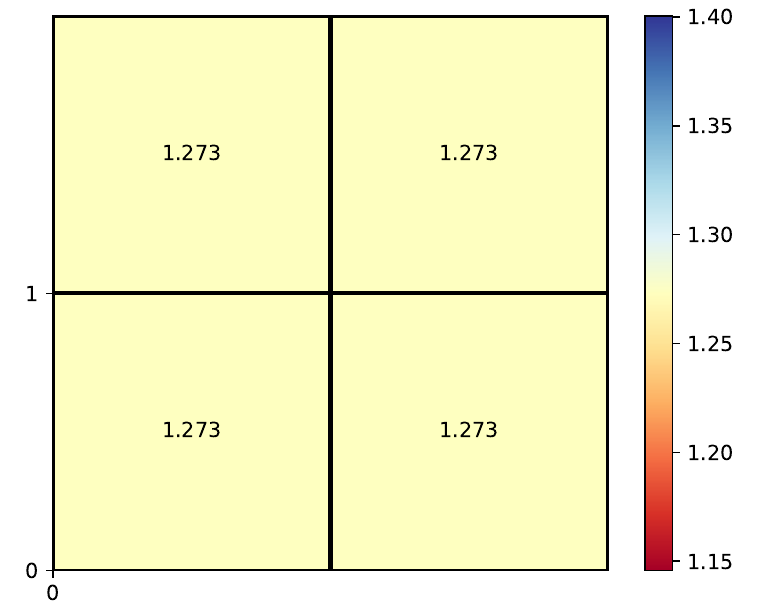}\\(b)}
\end{minipage}
\caption{The output of the ``Shape factor'' method: (a) visualization of particles and (b) Heatmap for the parameter $\vartheta_{jLS}$.}
\label{fig:ShapeCoefficientParticleVisualizationHeatmap}
\end{figure}

The user provides a \emph{.txt} file containing the particle coordinates. There are two options available: either upload an existing file, or generate a new one automatically, while specifying the number of particles to be generated. For example, the user provides an existing file located in \textit{ISANM} folder as input. Next, the user needs to manually assign the status of the primary particle. For example, the primary particle will be the isolated S particle. After clicking the ``Run'' button, script starts processing the input data and subsequently provides the results (Fig.~\ref{fig:ShapeCoefficientParticleVisualizationHeatmap}), which can later be saved.

\section{System Overview}

\subsection{System Architecture}

The system, accessible at \href{https://isanm.space}{https://isanm.space}, employs a client-server architecture. Users interact with the system via a web browser and authenticate to access its features. The server manages incoming requests, performing the required operations and accessing the database as needed. Once the request is processed, the server generates a response and sends it back to the user. \emph{Python} scripts are integrated into the \emph{API} backend, processing input data with the help of imported libraries to produce the desired output.

\begin{figure}[h]
\center{\includegraphics[width=0.95\linewidth]{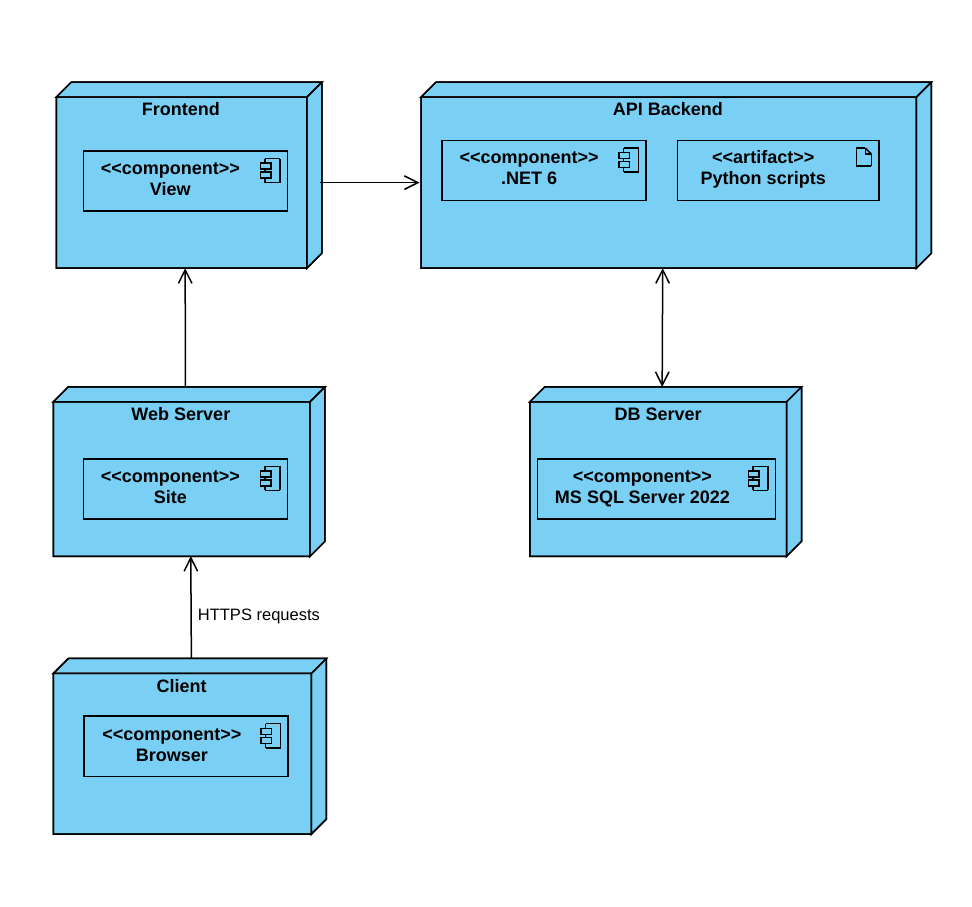}}
\caption{System architecture.}
\label{fig:SystemArchitecture}
\end{figure}

The system has been developed using the \emph{C\#} programming language and the \emph{Microsoft .NET 6} platform. It is designed to efficiently manage and process data while ensuring robust functionality and user interaction. For effective data storage and handling, the system employs the \emph{MS SQL Server 2022} as its database management system. A key advantage of the system is its use of \emph{Microsoft .NET 6}, which provides a powerful and flexible framework that enables rapid deployment and scalability. The complete architecture of the system is illustrated in Fig.~\ref{fig:SystemArchitecture}.

\subsection{System deployment}

The system has been deployed on a VDS server running the Linux operating system, specifically the Ubuntu 20.04 distribution. The server configuration includes 2 CPUs at 3.3 GHz, 4 GB of RAM, and 50 GB of NVMe storage. The web server in use is Nginx, which manages incoming requests and forwards them to the backend application built on .NET 6, running on a separate port. This architecture ensures a clear separation of responsibilities between the web server and the application, simplifying SSL certificate management for establishing a secure connection. The database management system employed is MS SQL Server, providing an efficient and reliable solution for data storage. The system is accessible at \href{https://isanm.space}{https://isanm.space}.

\subsection{System functionality}

\begin{figure}[h]
\center{\includegraphics[width=0.95\linewidth]{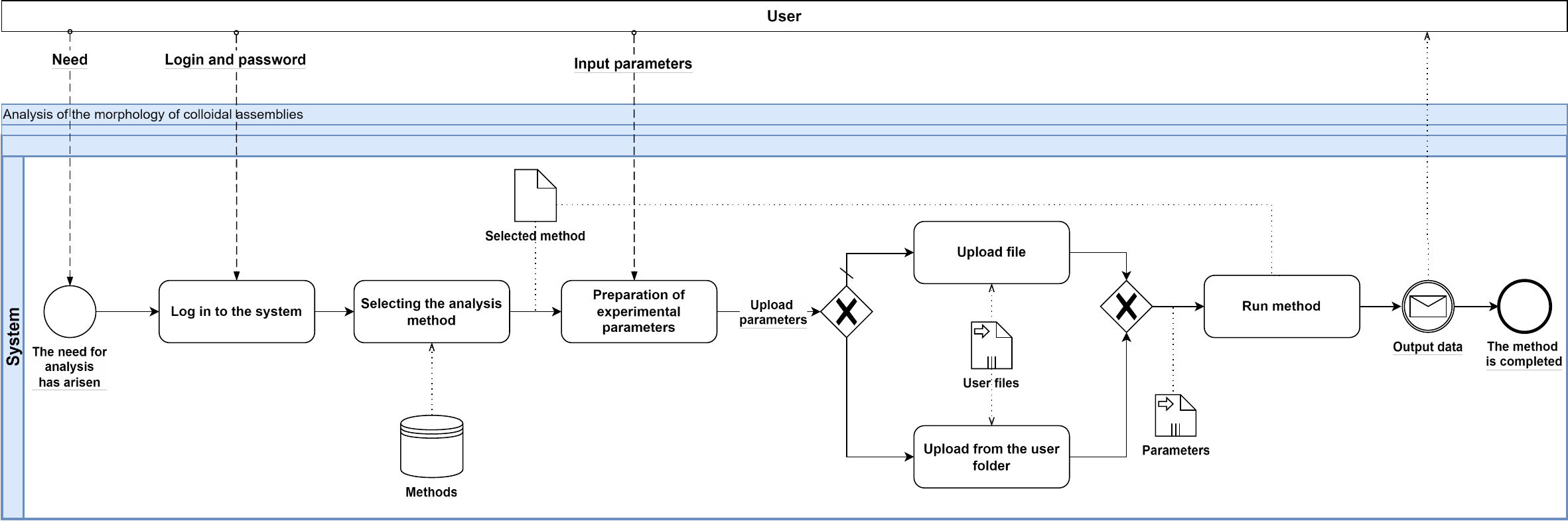}}
\caption{\emph{Workflow} diagram.}
\label{fig:WorkflowDiagram}
\end{figure}

The system allows users to select from various methods for analyzing colloidal assembly morphology and guides them through all stages, from importing input parameters to obtaining results. This section provides an overview of a typical system workflow and describes its core functions. The workflow, illustrated in BPMN notation, is shown in Fig.~\ref{fig:WorkflowDiagram}.

\subsection{User authorization in the system}

\begin{figure}[h]
\center{\includegraphics[width=0.95\linewidth]{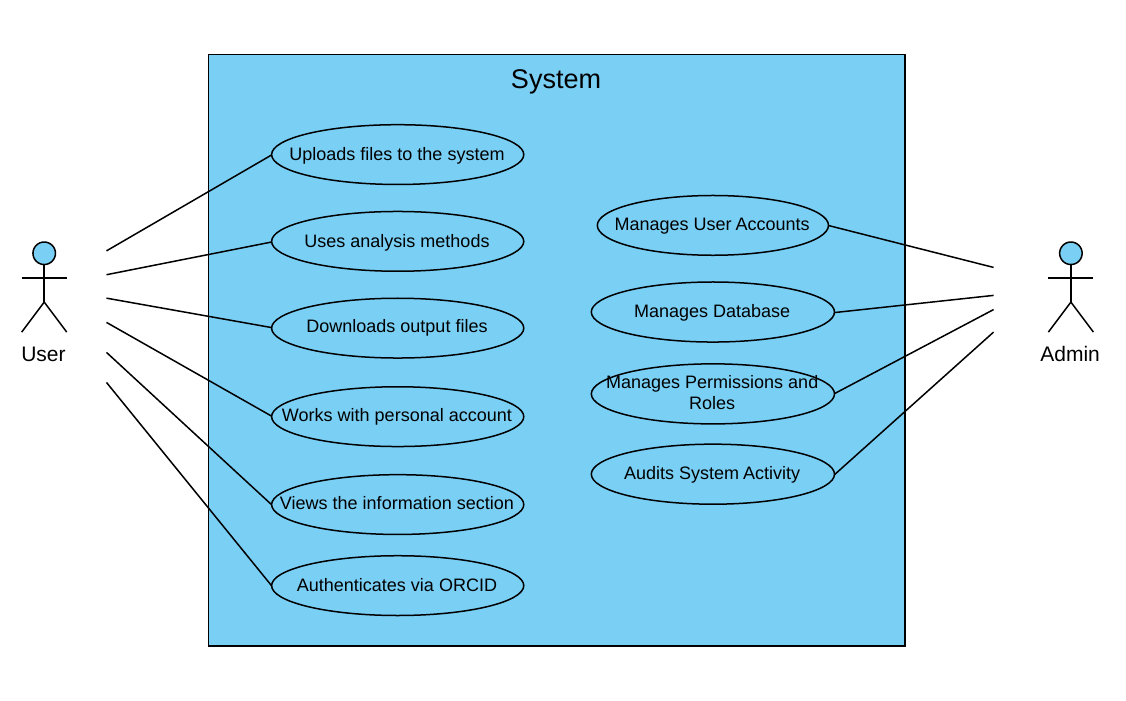}}
\caption{\emph{UseCase} diagram.}
\label{fig:UseCase}
\end{figure}

When the system is launched, users are immediately directed to the authentication page, where they need to complete the authentication process to access the full range of functionalities. The available features will vary according to the authenticated user’s role. Currently, users can authenticate using their Open Researcher and Contributor ID (\emph{ORCID}), allowing users with an existing \emph{ORCID} account to log into the system seamlessly. Authentication is performed via the \emph{ORCID API} (see \href{https://orcid.org/}{https://orcid.org/}), which provides secure access to user data. This method not only streamlines the registration and login process but also enables users to leverage their existing accounts. As a result, this approach safeguards the system from unverified users while also simplifying and automating the authentication process. The full range of system functionalities available to users is depicted in the \emph{Use Case} diagram shown in Fig.~\ref{fig:UseCase}. If desired, after logging in for the first time, one can confirm own email address and create a username/ password to log into their personal account.

\subsection{Selection of analysis method}

After successful authentication, users are presented with a range of methods for analyzing the morphology of colloidal assemblies. These methods are retrieved from the database and shown in a dropdown list on the analysis methods page. Currently, the system offers a total of nine methods: the orientational order parameter, radial uniformity, shape factor, bond length, mean squared displacement, Voronoi diagram, concentric tetratic order parameter, nematic order parameter, and a module for calculating statistical errors. Each method is thoroughly described on the system information page. Future plans include the implementation of additional methods.

\subsection{Preparation of experimental data}

\begin{figure}[h]
\center{\includegraphics[width=0.95\linewidth]{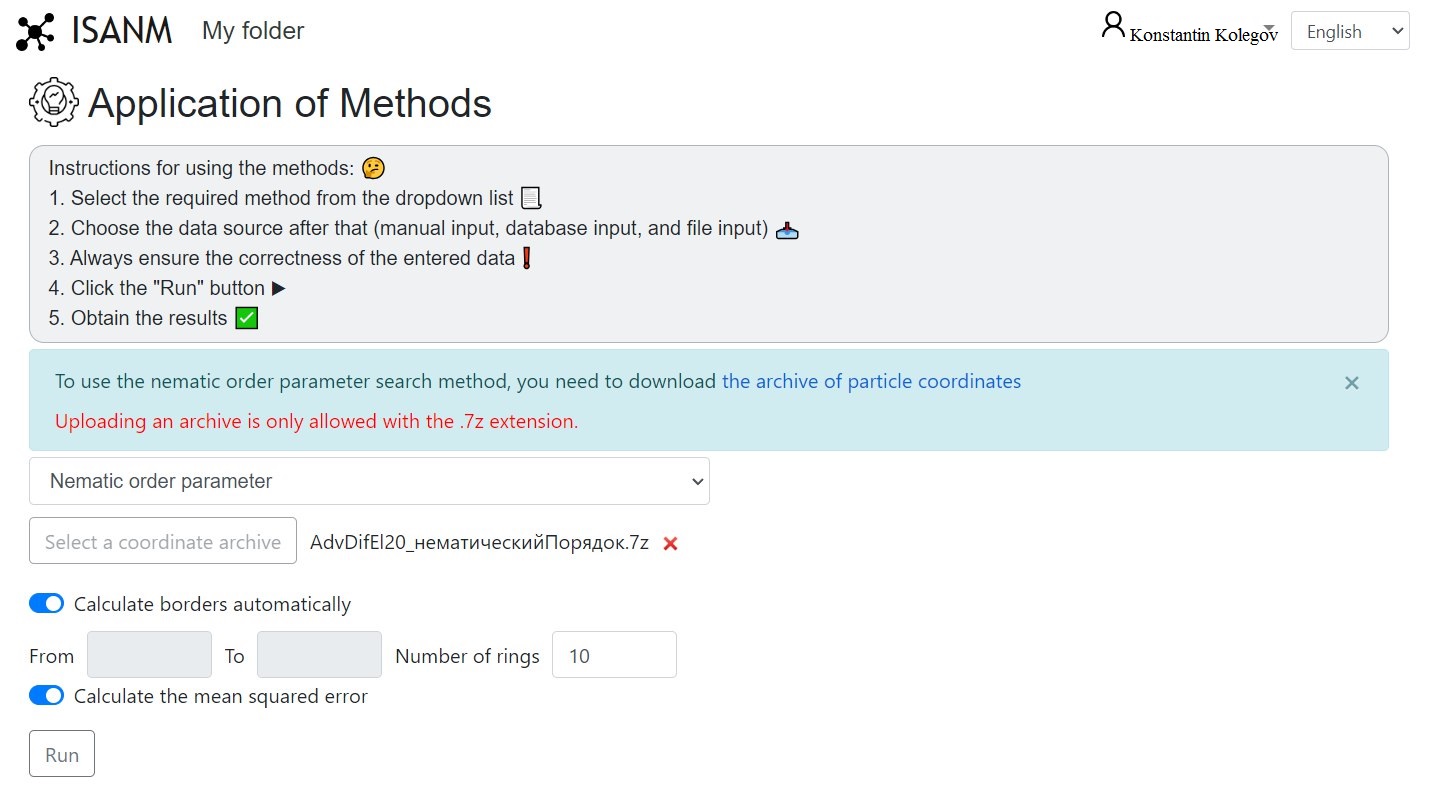}}
\caption{Example of method functionality.}
\label{fig:ExampleOfMethod}
\end{figure}

Most methods in the system require particle coordinates as input, and data can be imported in two ways: either through automatic generation of particle coordinates or by uploading files from the user folder. The server, built on \emph{.NET}, along with \emph{Python} scripts, reads and validates the data sets. Currently, the following data formats are used: .7z archives containing experiment folders with text files of particle coordinates, .txt files with particle coordinates at specific time steps, and raster image files of colloidal assemblies. An example of the parameters page is shown in Fig.~\ref{fig:ExampleOfMethod}.

\subsection{Method Launch}

After the input data has been uploaded and validated, the user can initiate the method launch. During this stage, the input parameters and their related information are accessed by the .NET server and sent to a Python script for further execution using the \emph{ProcessStartInfo} class. The output generated is then retrieved by the .NET server via the \emph{ProcessStartInfo} class before being presented to the user.

\section{Impact}

The proposed system integrates various methods for analyzing colloidal structures in one platform. This online service is beneficial for engineers and scientists working with colloidal assemblies and their applications. In addition, this system is useful for teachers and students, as it can be integrated into the learning process. Using the \emph{ISANM} information system for data processing allows users to automate workflows without the need for programming skills. This will speed up research and improve the quality of education for chemical engineers and beyond.

Characterizing morphology, identifying patterns, and quantitatively assessing order within colloidal assemblies are important for several reasons. First, it is beneficial to compare various self-assembly methods and evaluate the impact of different process parameters on the final colloidal structure. Additionally, studying such structures formed by colloidal particles can enhance our understanding of particle interactions and the phase transitions. Finally, the growing interest in colloidal structures within materials science for practical applications~--- ranging from optoelectronics to biosensors~--- necessitates a thorough characterization of colloidal structure morphology due to its close correlation with the physical properties of materials~\cite{Lotito2019}. Developing tools for analyzing the morphology of colloidal assemblies has a significant impacts on ongoing research in colloidal chemistry and nanotechnology. Currently, \emph{ISANM} includes only a basic set of implemented methods, with plans for future enhancements and expansions of the information system.

\section{Further work}
In the future, it is planned to increase the number of analysis methods in the system, including those related to image processing and based on machine learning and neural networks. The reference materials can be made even more visual and user-friendly, including new methods. It is also planned to introduce a number of mathematical models describing the self-assembly of colloidal structures. Another important area is the integration of work with other tools, such as \emph{ImageJ} (data export and import, similar data formats, and so on). It may be useful to be able to share the user's folder via a link or another method. This will allow students to carry out joint projects. They will have convenient opportunities to demonstrate their results to teachers.  Teachers will be able to share their materials with students. Researchers will organize collaboration in their work. Feedback will definitely be implemented, which will allow us to make timely edits to the system and add new functionality based on user wishes.

\section*{Conclusion}

The developed Information system for analysis of nanostructure morphology, \emph{ISANM}, provides experts in the relevant field with useful tools for their work, requiring no programming skills. This system functions as a client-server application, ensuring ease of use, accessibility, and the delegation of computational tasks from the client to the server. Users can upload and store their files in the system, categorize them into thematic folders, apply various processing methods, save the results, and download output files containing the processed data. In the future, the developed tool will be useful for scientists and engineers working in the fields of colloidal chemistry, materials science and nanotechnology, which will affect the increase in labor productivity. In our opinion, introducing this system into the educational process would improve the quality of training for future specialists.

\section*{Acknowledgement}
This work is supported by Grant No. 22-79-10216 from the Russian Science Foundation (\href{https://rscf.ru/en/project/22-79-10216/}{https://rscf.ru/en/project/22-79-10216/}).

 \appendix

\section{Integration of Python scripts into a .NET application}
\label{appendix:PythonIntegration}

An essential feature of the system is the integration of \emph{Python} scripts within the \emph{.NET} application. In this context, the \emph{ProcessStartInfo} class from the \emph{System.Diagnostics} namespace is employed, which not only sets up the properties of a process prior to its launch but also facilitates its management. Executing \emph{Python} scripts is crucial during the method selection and the transmission of input parameters. Listing~\ref{lst:PythonIntegration} illustrates how one of the methods is executed and handled using the \emph{ProcessStartInfo} class.

\begin{lstlisting}[language={[Sharp]C}, caption={Running a Python script in a .NET application}, label={lst:PythonIntegration}]
// for nematic order parameter
string zip = splitted[1]; // link to input data archive
string checkError = splitted[2]; // whether to calculate the error or not
string selectedError = splitted[3]; // chosen error
// generate random name for the extracting folder
Random rand = new Random();
var folderName = GetRandomString();
var targetFolder = @"wwwroot/" + folderName;
// get archive extension
var extArchive = zip.Substring(zip.LastIndexOf("."));
// check for archive extension
if (extArchive.Equals(".7z")){
    var left = splitted[4]; // left border parameter
    var right = splitted[5]; // right border parameter
    var Nring = splitted[6]; // number of rings
    var autoBorders = splitted[7]; // automatic calculation of particles boundaries
    var webroot = this.Environment.WebRootPath; // root directory
    // extracting the input archive to the folder
    using (SevenZipArchive archive = new SevenZipArchive(zip)){
        archive.ExtractToDirectory(targetFolder);
    }
    var psi = new ProcessStartInfo();
    psi.FileName = @"python.exe";
    string scriptt = Path.Combine(webroot, "methods", "nematicParameter.py"); // path to script file
    psi.Arguments = $"\"{scriptt}\" \"{webroot + "\\" + folderName}\" \"{checkError}\" \"{selectedError}\" \"{left}\" \"{right}\" \"{Nring}\" \"{webroot}\" \"{autoBorders}\""; // passing the process parameters
    psi.UseShellExecute = false; // launch the process directly
    psi.CreateNoWindow = true; // do not open additional window
    psi.RedirectStandardOutput = true; // read the process output through the StandardOutput
    psi.RedirectStandardError = true; // read the process errors thought the StandardError
    // start method
    using (var process = Process.Start(psi)){
        errors = process.StandardError.ReadToEnd(); // get obtainer errors
        result = process.StandardOutput.ReadToEnd(); // get obtainer result
    }
    // python script returns the name of the generated file
    var fileName = result; // new file name rS
    // archive obtained files
    if (checkError == "true") // if you select to plot a graph, then save two files
    {
        // create FileStream for the output ZIP archive
        using (FileStream zipFile = System.IO.File.Open("outputS.zip", FileMode.Create))
        {
            // file added in archive
            using (FileStream source1 = System.IO.File.Open("rS.txt", FileMode.Open, FileAccess.Read))
            {
            // second file added in archive
                using (FileStream source2 = System.IO.File.Open(chooseError == "std_MSE" ? "MSE.txt" : "SEM.txt", FileMode.Open, FileAccess.Read))
                {
                    using (var archive = new Archive())
                    {
                        // add all files in archive
                        archive.CreateEntry("rS.txt", source1);
                        archive.CreateEntry(chooseError == "std_MSE" ? "MSE.txt" : "SEM.txt", source2);
                        // final arciving
                        archive.Save(zipFile, new ArchiveSaveOptions() { Encoding = Encoding.ASCII, ArchiveComment = "two files are compressed in this archive" });
                    }
                }
            }
        }
    }
}
\end{lstlisting}

Initially, input parameters from the client side are initialized. These vary by method. For example, the nematic order parameter method uses a coordinate archive, which is extracted on the server, and the system currently supports only .7z format.
The \emph{ProcessStartInfo} class is initialized in the code, and the necessary parameters are passed to it for execution. To initiate each process, the path to the Python script is first specified, followed by the input parameters for the method. Errors and results are captured in the ``errors'' and ``result'' variables, helping monitor execution. The output is a path (or paths) to files or plots generated by the script.

  \bibliographystyle{elsarticle-num}
  \bibliography{Khusainova2024}





\end{document}